\documentclass[runningheads]{llncs}

\usepackage{graphicx}
\usepackage{upquote}
\usepackage{booktabs}
\usepackage{url}
\usepackage{xcolor}
\usepackage{listings}
\usepackage{chngcntr}
\usepackage{pdfcomment}
\usepackage{hyperref}
\usepackage{paralist}
\usepackage[math]{blindtext}
\usepackage{mwe}
\usepackage{comment}
\usepackage{tikz}
\usepackage{subfigure}
\usetikzlibrary{backgrounds,automata}
\usepackage{multirow}
\usepackage{booktabs}
\usepackage{array}
\usepackage{amsmath}
\usepackage[capitalise,nameinlink]{cleveref}

\usepackage{lineno}
\usepackage{pgfplots}
\lstdefinelanguage{rebeca}{
  morekeywords={reactiveclass, knownrebecs, statevars, main, msgsrv, main, define, LTL, CTL, boolean, int, shortint, byte, if, else, while, for, wait, msg, reset, set, self, false, true, now, after, delay, deadline, initial, env},
  otherkeywords={=>,<-,<\%,<:,>:,\#,@},
  sensitive=true,
  morecomment=[l]{//},
  morecomment=[n]{/*}{*/},
  morestring=[b]",
  morestring=[b]',
  morestring=[b]"""
}

\usetikzlibrary{patterns}

\begin{document}
\title{Schedulability Analysis of WSAN Applications:\\ Outperformance of A Model Checking Approach}
\titlerunning{Where Model Checking Outperforms Analytical Approaches}
%
%

\author{Ehsan Khamespanah $^{1,2}$ \and 
Morteza Mohaqeqi \and \\ Mohammad Ashjaei $^{3}$ \and Marjan Sirjani $^{2,3}$}
\institute{$^1$ School of Electrical \& Computer Engineering, University of Tehran, Iran \\ $^2$ School of Computer Science, Reykjavik University, Iceland \\ $^3$ School of Innovation, Design, and Engineering,  M{\"a}lardalen University, Sweden}

\authorrunning{Khamespanah et al.}

\maketitle              
\begin{abstract}
Wireless sensor and actuator networks (WSAN) are real-time systems which demand high degrees of reliability requirements. To ensure this level of reliability, different analysis approaches have been proposed for WSAN applications. Among different alternatives, analytical analysis and model checking are two common approaches which are widely used for the formal analysis of WSAN applications. Analytical approaches apply constraint satisfaction methods, whereas model checking generates explicit states of models and analyze them. In this paper, we compare the two approaches in schedulability analysis of WSAN applications using an application for monitoring and control of civil infrastructures, which is implemented on the Imote2 wireless sensor platform. 
We show how the highest possible data acquisition frequency for this application is computed while meeting the deadlines, and compare the results of the two approaches as well as their scalability, extensibility, and flexibility.

\end{abstract}

\section{Introduction}
Wireless sensor and actuator networks (WSANs) provide low-cost continuous monitoring as the infrastructure for control systems. Thanks to the use of wireless communications and distributed architectures, WSANs encompass many advantages as compared to traditional hard-wired networked control systems. In the design of control systems which benefit from WSANs, interdependencies between control algorithm and communication have to be considered. For example, while low sampling rate for sensor data acquisition usually degrades control performance, high sampling rate may increase resource contention in bandwidth-constrained WSANs or high power consumption, which may lead to degrading the control performance too. The coupling between wireless communication and control algorithm, therefore, motivates using cyber-physical co-design approach that integrates wireless networks and control systems designs. 

This makes WSAN applications design more complicated and it is necessary to develop techniques and methods that facilitate designing and verification of WSAN applications. However, trial and error is the widely used approach for this purpose. For example, in~\cite{Linderman2012}, an empirical test-and-measure approach based on binary search is used to find configuration parameters, including worst-case task runtimes, and time-slot length of the communication protocols. 
Extending real-time scheduling theory~\cite{sha2004real} that has been developed for the analysis of real-time systems is another approach for the analysis of WSAN environments. Unlike a real-time operating system (RTOS), the event-driven operating systems of WSAN applications (e.g. TinyOS~\cite{tinyos}) generally do not provide real-time scheduling guarantees, priority-based scheduling, or resource reservation functionality. Without such support, many schedulability analysis approaches cannot be effectively employed.

On the other hand, a set of techniques have been proposed to provide schedulability analysis, including utilization-based tests~\cite{Liu_JACM73}, response-time analysis~\cite{Joseph_CJ86}, and real-time calculus~\cite{Thiele_IEEE00} as samples of analytical approaches and model checking approaches~\cite{Waszniowski2008,Yi_tool04}. Considering the analytical approaches, several works in the literature addressed the schedulability analysis of WSANs considering different settings and requirements. Authors in~\cite{Xu_GCC13} defined three different task scheduling techniques, then presented an analytical method to find the maximum schedulable load. The work in~\cite{Xia_IEEE17} and~\cite{Cheng_TOR11} presented analytical analysis;  in~\cite{Xia_IEEE17} the target is the earliest deadline first scheduling algorithm, and in \cite{Cheng_TOR11} the target is the high energy first scheduling algorithm.

Among the works that analyzed real-time systems using model checking approaches, 
Petri Nets are used to check the behavior of task executions in \cite{Furfaro_MCSIT07,Xu_IEEE02}. 
In the context of uniprocessor real-time systems, the authors in~\cite{Fersman_tools02} developed a timed automata model for schedulability analysis of preemptable tasks that have interactions with each other through synchronization methods. A technique in~\cite{Li_IPDPSW10} is proposed to tackle similar problem in analyzing tasks executing synchronously. The work in~\cite{Fersman_JC04} extended the reachability analysis on timed automata to consider precedence and resource constraints in uniprocessor real-time systems.
In the domain of multiprocessor systems, the authors in~\cite{Sheng_WMTV10} proposed a schedulability analysis based on model checking for multiprocessor real-time systems. 
The authors in~\cite{Guan_ISORC08} perform exact schedulability analysis of multi-core real-time systems using model checking.

To the best of our knowledge, although each of the above mentioned techniques widely used 
in the analysis of WSAN applications, it is not clear which of them has to be used for the analysis of a given WSAN application. In other words, there is no work on comparing analytical and model checking approaches to make clear their strengths and weaknesses against each other.
The main reason is that the techniques are normally optimized for specific applications, which makes it difficult to globally compare the techniques. 
In this paper we compare analytical approach 
and model checking approach in the analysis of a WSAN framework which is proposed for developing structural health monitoring and control applications~\cite{Linderman2012}. 
This framework has been implemented on the Imote2 wireless sensor platform and used in several long-term developments of highway and railroad bridges. Note that applications of this framework can be extended to a wide range of monitoring purposes as it has a flexible and open architecture~\cite{spencer2015}. 
%
To have a better understanding of this framework, in Section~\ref{sec::WSAN} we present the specification of it in a sequence diagram. 
We show the analytical analysis of this framework in Section~\ref{sec::analytical}, and the model checking approach in Section~\ref{sec::modelchecking}.

The results of our study, presented in Section~\ref{sec::comp}, show that the analytical analysis for the application described in this paper which is based on conservative worst-case assumptions and empirical measurements,  may lead into schedules that do not utilize resources in the most efficient way. By using model checking we can get better utilization of resources. However, model checking is more sensitive to the values which are set for timings of the model and different values may result in different memory and time consumption for model checking the models. Note that having different values for timings do not affect the analysis cost of the analytical approach.
We also compare the extensibility and flexibility of the two approaches by checking how each approach can be reused for different network communication protocols and different characteristics of tasks.

\section{Application Model and Design Objectives} \label{sec::WSAN}

WSANs provide a new communication technology for a wide range of cyber-physical applications. A WSAN involves feedback control loops between sensors and actuators through a wireless network. Sensors of WSANs measure process variables and deliver the set of gathered data to a controller through the network. The controller sends control commands to the actuators, which then operate the control and safety components. Structural health monitoring and control (SHMC) of civil infrastructure~\cite{Linderman2012} is an instance of WSANs, used in several long-term development of several highway and railroad bridges~\cite{spencer2015}. SHMC application development has proven to be particularly challenging: it has the complexity of a large-scale distributed system with real-time requirements, while having the resource limitations of low-power embedded WSAN platforms. Sensing and control in these systems need to meet stringent real-time performance requirements on communication latency in harsh environments. Violation of these requirements may result system failure or significant costs.

A WSAN application is a distributed system with multiple sensor nodes, each comprised of the independent concurrent entities: a CPU and some sensors which bridged together via a wireless communication device which uses a transmission control protocol. Interactions between entities, both within a node and across nodes, are concurrent and asynchronous. Moreover, WSAN applications are sensitive to the timing of events, with soft deadlines at each step of the process that are required to ensure correct and efficient operation. Due to the performance requirements, coordination among sensing, data processing, and communication activities is required in WSANs. In particular, once a sample is acquired from a sensor, its corresponding radio transmission activities must be performed. At the same time, data processing tasks must be executed (for example, because of the environmental changes in the temperature, a kind of data compensation must be applied on sensor data to adjust the acquired values). Moreover, the timing of radio transmissions from different nodes must be coordinated using a communication protocol.


\subsection{The System Model of a WSAN for Structural Health Monitoring and Control} 
\begin{figure*}
\centering
\includegraphics[width=1.0\textwidth]{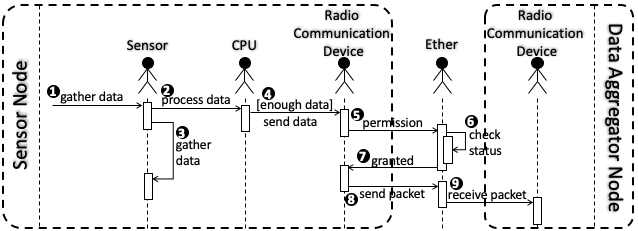}
\caption{The sequence diagram of the main scenario of a WSAN for Structural Health Monitoring and Control}
\label{fig::WSAN-seq-diagram}
\end{figure*}

As shown in Figure~\ref{fig::WSAN-seq-diagram}, we use a UML sequence diagram to present how WSAN applications work. To this end, we need to have a look into the interaction of its components and the events which are triggered and served by them. Using this diagram, we developed two approaches (presented in Section~\ref{sec::analytical} and \ref{sec::modelchecking}) for the purpose of checking for meeting the analysis objectives of the model. 

Based on the specification of the WSAN applications, there are many nodes which have the role of data acquisition and data transmission. 
Note that all of the message calls in this model are asynchronous; so, they are shown by open arrows in Figure~\ref{fig::WSAN-seq-diagram} and they do not have any return value. 

For data acquisition, a node has a set of sensors which periodically acquire data from the environment (messages 1 and 3 in Figure~\ref{fig::WSAN-seq-diagram}) and send the data to the processing unit of the node (message 2 in Figure~\ref{fig::WSAN-seq-diagram}). The processing unit is responsible for validating the data and storing it into an internal buffer. Upon receiving enough data (the guard on message 4 in Figure~\ref{fig::WSAN-seq-diagram}), the processor unit puts the data in one packet and sends it to the radio communication unit (message 4 in Figure~\ref{fig::WSAN-seq-diagram}). The radio communication unit tries to send data via a wireless medium (message 5 in Figure~\ref{fig::WSAN-seq-diagram}), considering a predefined communication protocol (message 6 to 9 in Figure~\ref{fig::WSAN-seq-diagram}). In this work, we assumed that the communication protocol of nodes is TDMA.  Note that TDMA~\cite{DBLP:conf/iscc/El-Hoiydi02} is a MAC-level communication protocol which is widely used in WSAN applications. But it can be replaced by other MAC-level protocols (e.g. B-MAC~\cite{DBLP:conf/sensys/PolastreHC04}). In addition to the mentioned components, one additional component for carrying out miscellaneous tasks unrelated to sensing or communication is needed in the model of WSAN application. This additional component is necessary for making the model close to its real configuration. 

Regarding an implementation point of view, each node employs a single-core processing unit. The mentioned functionalities are assumed to be implemented by two real-time tasks, referred to as sensor task and miscellaneous task. Further, the corresponding jobs are scheduled and executed by a non-preemptive FIFO scheduling policy.

\subsection{Design Objective}
In the 
WSAN application the goal is that each node reaches its maximum sampling rate.
This way the resource is fully utilized and the cost is reduced. Meanwhile, system buffer size as well as its processing capacity are the main limitations. 
In summary, the problem is to compute the maximum achievable 
sampling rate of sensors for a given system specification
without missing any data.

The constraints which have to be satisfied in finding the maximum sampling rate can be divided into two requirements. The first requirement focuses on the timely execution of the tasks inside a node, which is expressed in Requirement~\ref{req::inter}.

\newtheorem{require}{Requirement}

\begin{require}
\label{req::inter}
All instances of sensors tasks as well as those of any other miscellaneous tasks should be served prior to the arrival of a new instance of that task.
\end{require}

The second requirement, Requirement~\ref{req::intra}, is related to the communication protocol parameters. Once a packet is ready in a node, this packet should be sent before another packet becomes ready for sending in that node. 
 
\begin{require}
\label{req::intra}
The transmission of any packet should be finished before the next packet becomes ready. 
\end{require}
 
\section{Analytical Analysis Approach} \label{sec::analytical}

In this section, we present an analytical approach to obtain the maximum sampling rate of data collection. Our solution provides 
a rate by which both requirements are met (but maybe it is not the maximum possible rate); hence, the solution is a \emph{safe} (but possibly pessimistic) approximation. Note that in this section we do not use schedulability analysis tools like chronVAL/chronSIM \cite{DBLP:conf/erts/AnssiKMS12} which provide facilities for modeling, but use the same technique for analyzing models and compute results with the same characteristics.

Based on the description of the previous section, a node contains two tasks, namely $\tau_M$ and $\tau_S$. Both tasks are assumed to be periodic with period of $T_M$ and $T_S$, although their activation sources are an event for $\tau_M$ and a timer for $\tau_S$. The tasks have a known worst-case computation time, which is denoted by $C_M$ for $\tau_M$ and $C_S$ for $\tau_S$. Whenever any of these tasks is activated, it is inserted to a FIFO queue to be scheduled non-preemptively. Each task has a relative deadline that is equal to its period. In other words, we consider the model to be \emph{implicit}  deadline. Note that the tasks do not have activation jitter and they have no dependency, such as sharing resources or communication between each other.

In this model, the Sensor task  $\tau_S$ collects data from the environment. When $N$ data samples are collected and processed by this task, a packet is created for sending the data through the network. 

\begin{figure*}[!htbp]
\centering
\begin{tikzpicture}[x=1.2cm,scale=0.6]
 Drawing the axis:
\draw[black,->,>=latex]
  (0,0) -- (16,0) node[below right] {$ t$};

\draw[black,->,>=latex]
  (0,-3) -- (16,-3) node[below right] {$ t$};


\draw[] (0,-3) -- ++(0,-5pt) node[below] {$0$};


\draw[->,thick,>=latex] (0,0) node[below] {$0$} --  (0, 1);
\draw[->,thick,>=latex] (2,0) node[below] {$T_S$} --  (2, 1);
\draw[->,thick,>=latex] (4,0) node[below] {$2T_S$} --  (4, 1);
\draw[->,thick,>=latex] (8,0) node[below] {$(N-1)T_S$} --  (8, 1);
\draw[] (6,0.3) -- (6,0.3) node        {$\ldots$};
\draw[] (12,0.3) -- (12,0.3) node        {$\ldots$};

\draw[pattern=dots] 			(0,0) rectangle (1, 0.5);
\draw[pattern=dots] 			(2,0) rectangle (3, 0.5);
\draw[pattern=dots] 			(4,0) rectangle (5, 0.5);
\draw[pattern=dots] 			(8,0) rectangle (9, 0.5);

\draw (10,-1.5) circle (2ex);

\draw[pattern=north west lines] (0.5,-3) rectangle (1.5, -2.5);
\draw[pattern=north west lines] (10.5,-3) rectangle (11.5, -2.5);

\draw[black,<->,>=latex]  (0.5,-3.5) -- (10.5,-3.5) node[midway,yshift=-0.4cm] {$T_\mathit{tdma}$};

\draw[->] (9,0)  --  (9.9 , -1.2) node[right,midway] {\scriptsize Preparing a packet};
\draw[->] (10,-1.8)  --  (10.5 , -2.5) node[right,midway] {\scriptsize Sending the packet};

\draw[pattern=dots]             (0,-5.5) rectangle (1, -5) node[right,yshift=-0.15cm] {\scriptsize Execution of $\tau_S$ (when ignoring $\tau_M$)};
\draw[pattern=north west lines] (0,-6.5) rectangle (1, -6) node[right,yshift=-0.15cm] {\scriptsize Dedicated TDMA time slot};

 \end{tikzpicture}
\caption{Creating a packet out of $N$ data samples. The packet is sent in the first dedicated TDMA time slot.}
\label{fig:tdma}
\end{figure*}
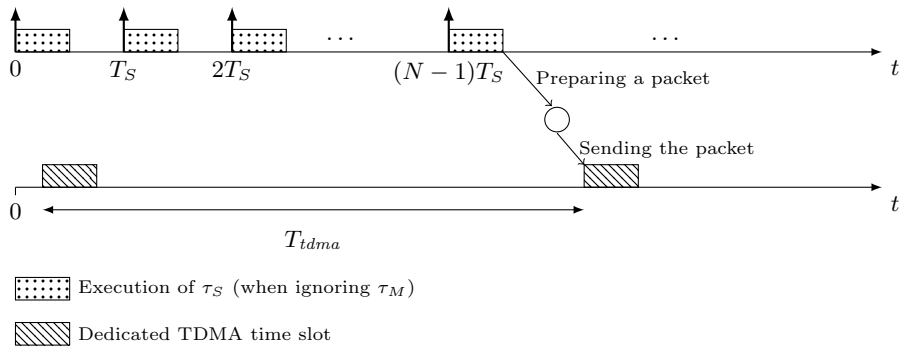

As mentioned before, the network access protocol is assumed to be TDMA and each sender node has a dedicated time-slot for transmission of its data. In this model, we define a super-frame with the length of $T_\mathit{tdma}$ that all time-slots are allocated within that. If the packet is not ready at the beginning of the dedicated time-slot, it will be unused.  Figure~\ref{fig:tdma} demonstrates the execution of $N$ jobs of task $\tau_S$, preparing a packet of the collected data, and sending the packet in the dedicated time slot.

Regarding the design objective, we are interested in finding the maximum allowable rate of activation of task $\tau_S$. A sampling rate is valid to be used if it respects system constraints, specified in Requirements~\ref{req::inter} and \ref{req::intra}. 
%
%
To treat the mentioned requirements, we define the notion of \emph{response time} of a job. Consider the $i$-th job of the Sensor task, that is the instance of $\tau_S$ which is responsible for processing the $i$-th data. The difference between the release time of this job and the point at which the data is completely processed is defined as the response time of the job and is denoted by $R_i$.

\subsection{Addressing Requirement~\ref{req::inter}}

To decide schedulability of the tasks, i.e., to fulfill Requirement~\ref{req::inter}, we need to ensure that all instances of both tasks meet their deadlines. In the following, we derive a relation which guarantees tasks schedulability. 

Each sender node has two periodic non-preemptive tasks which are scheduled with a FIFO queue. As there is no other priority levels in the system, the only interference for the tasks is from the same priority tasks waiting in the same queue. Therefore, in a schedulable system with implicit deadlines,  when a task is queued in the FIFO queue at most one instance of other tasks can be ahead of the task under analysis~\cite{Davis_11}. This means that if there are two instances of a task ahead of the task under analysis, the system is not schedulable. According to the assumptions, a task suffers from the worst-case computation time of the other task, only. Therefore, in our system model with two periodic non-preemptive tasks, the task $\tau_S$  is schedulable if $C_M + C_S \leq T_S$. Similarly, the task $\tau_M$ is schedulable if $C_M + C_S \leq T_M$. Putting these together, the system is schedulable if 

\begin{equation}
\label{eq:suffTest}
C_M + C_S \leq min(T_M, T_S)
\end{equation}

Condition~(\ref{eq:suffTest}) serves as a sufficient schedulability test, i.e., a sufficient condition for fulfilling Requirement~\ref{req::inter}.   







\subsection{Addressing Requirement~\ref{req::intra}}
\begin{figure*}[!htbp]
\centering
\begin{tikzpicture}[x=1.2cm,scale=0.6]
 Drawing the axis:
\draw[black,->,>=latex]
  (0,0) -- (17,0) node[below right] {$ t$};

\draw[->,thick,>=latex] (0,0) node[below] {\tiny $0$} --  (0, 1);
\draw[->,thick,>=latex] (1.2,0) node[below] {\tiny $T_S$} --  (1.2, 1);
\draw[->,thick,>=latex] (2.4,0) node[below] {\tiny $2T_S$} --  (2.4, 1);
\draw[->,thick,>=latex] (4.8,0) node[below] {\tiny $(N-1)T_S$} --  (4.8, 1);
\draw[->,thick,>=latex] (6,0) node[below] {\tiny $NT_S$} --  (6, 1);
\draw[->,thick,>=latex] (11,0) node[below] {\tiny $((j-1)N)T_S$} --  (11, 1);
\draw[->,thick,>=latex] (15.8,0) node[below] {\tiny $(jN-1)T_S$} --  (15.8, 1);

\draw[] (4,0.5)  -- (4,0.5)  node        {$\ldots$};
\draw[] (9,0.5)  -- (9,0.5)  node        {$\ldots$};
\draw[] (14,0.5) -- (14,0.5) node        {$\ldots$};

\draw[pattern=dots] 			(0,0)    rectangle (0.5, 0.5);
\draw[pattern=dots] 			(1.2,0)  rectangle (1.7, 0.5);
\draw[pattern=dots] 			(2.4,0)  rectangle (2.9, 0.5);
\draw[pattern=dots] 			(4.8,0)  rectangle (5.3, 0.5);
\draw[pattern=dots] 			(6,  0)  rectangle (6.5, 0.5);
\draw[pattern=dots] 			(11,0)   rectangle (11.5,0.5);
\draw[pattern=dots] 			(15.8,0) rectangle (16.3,0.5);


\draw [decorate,decoration={brace,amplitude=10pt,mirror}]
		(0,-0.6) -- (5.5,-0.6) node [black,midway,yshift=-0.7cm] {\scriptsize First packet};
\draw [decorate,decoration={brace,amplitude=10pt,mirror}]
		(11,-0.6) -- (16.5,-0.6) node [black,midway,yshift=-0.7cm] {\scriptsize $j$-th packet };

\draw[->] (16.3,1.5)  node[above] {\scriptsize $m_j$} --  (16.3 , 0.5) ;

\draw[pattern=dots] (0,-3) rectangle (1, -2.5) node[right,yshift=-0.2cm] {\scriptsize Execution of $\tau_S$ (when ignoring $\tau_M$)};

 \end{tikzpicture}
\caption{The execution of any $N$ jobs of $\tau_S$ results a packet.}
\label{fig:m_j}
\end{figure*}
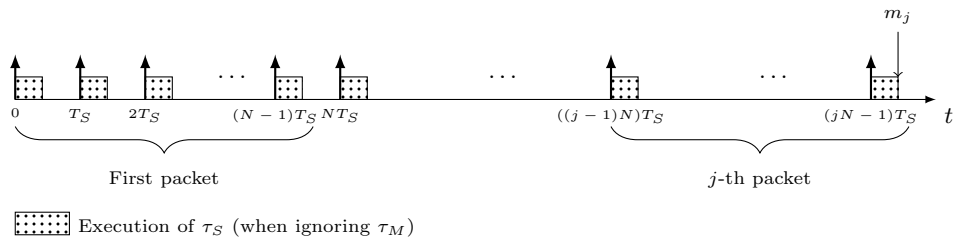

In order to address the second requirement, we first compute the time when the $j$-th packet becomes ready. Let $m_j$ denote this time instant. Figure~\ref{fig:m_j} depicts the relation between $m_j$ and the execution of $\tau_S$'s jobs. 

According to the figure, $m_j$ can be computed as
\begin{equation}
\label{eq:mj}
m_j = (j N-1) T_S + R_{j N},
\end{equation}
where $R_{j N}$ denotes the response time of the $jN$-th job of $\tau_S$ (note that the $jN$-th job is released at $(jN-1)T_S$ time instant).

In order for a packet to be sent before the next one is ready, there must be at least one dedicated time slot between the points in which the packets are ready. 
Formally, between the instants $m_j$ and $m_{j+1}$, for all $j>0$, there must be a time slot in which the $j$-th packet can be sent.

To fulfill this requirement, it is sufficient to have $T_\mathit{tdma} < m_{j+1}-m_{j}$, which, according to Eq.~(\ref{eq:mj}), means: 
\begin{eqnarray}
\label{eq:sched_condition}
T_\mathit{tdma} & < & m_{j+1}-m_{j}  \\
                & = & ((j+1)N-1) T_S + R_{(j+1) N} - \left( (j N-1) T_S + R_{j N} \right)  \nonumber \\
                & = & ((j+1)N) T_S   + R_{(j+1) N} -  (j N) T_S   - R_{j N}  \nonumber \\
                & = & N T_S  + R_{(j+1) N} - R_{j N}  \nonumber
\end{eqnarray}

This inequality forces a lower bound on the sampling period $T_S$ as follow:
\begin{equation}
\label{eq:lower-bound}
T_S >  \frac{T_\mathit{tdma}+R_{j  N}-R_{(j+1)  N}}{N}
\end{equation}
Hence, for an exact timing analysis (for the second problem), we need to precisely calculate the response times of the Sensor jobs.




\subsection{Response-Time Analysis}
As shown above, to compute the value of $m_j$, we need to accomplish a response-time analysis of a job that is scheduled by a non-preemptive FIFO scheduling scheme. As exact response-time of the jobs are not known, we use upper bound and lower bound of the response time to derive a (sufficient) condition which satisfies (\ref{eq:lower-bound}). 



\begin{proposition} \label{prop:imp}
Let $W$ and $B$ denote, respectively, an upper bound and lower bound for the response time of a job of $\tau_S$. Then, 
\begin{equation}
\label{eq:suff_condition}
T_\mathit{tdma}  <  N T_S + B -  W
\end{equation}
implies Eq.~(\ref{eq:sched_condition}).
\end{proposition}
\begin{proof}
Based on the definition of W and B, it holds that
\begin{eqnarray}
B         & \leq &  R_{(j+1) N},   \\
R_{j  N}  & \leq &  W, \nonumber
\end{eqnarray}
or equivalently,  
\begin{eqnarray}
 B  & \leq &  R_{(j+1) N}  \\
 -W & \leq &  -R_{j N}. \nonumber
\end{eqnarray}
This leads to 
\begin{equation}
\label{eq:BWRj}
B-W \leq R_{(j+1) N}-R_{j N}.
\end{equation}
With adding $NT_S$ to both left-hand side and right-hand side of (\ref{eq:BWRj}),
it follows that $N T_S +B-W \leq NT_S+R_{(j+1) N}-R_{j N}$. Writing this together with~(\ref{eq:suff_condition}), we have
\begin{eqnarray}
T_\mathit{tdma}  < & N T_S + B-W & \nonumber\\
                   &  N T_S + B-W & \leq NT_S+R_{(j+1) N}-R_{j N} \nonumber
\end{eqnarray}
This means that $T_\mathit{tdma} <N T_S+B-W$ implies $T_\mathit{tdma} < N T_s +R_{(j+1) N}-R_{j N}$, by which the proof completes.
\end{proof}

In other words, Eq.~(\ref{eq:suff_condition}) provides a sufficient condition to ensure that the messages are sent in a timely manner. Next, we consider calculating the parameters $B$ and $W$. The (minimum) execution time of a single job of the Sensor task can be considered as (a lower bound on) $B$. Besides, an upper bound for the response time of a job is $W=C_S+C_M$. Substituting these values of $B$ and $W$ in Eq.~(\ref{eq:suff_condition}) reveals:
\begin{equation}
\label{eq:suff_condition2}
T_\mathit{tdma}  <  N  T_S - C_M.
\end{equation}

\subsection{Maximum Sampling Rate}

Eqs.~(\ref{eq:suffTest}) and~(\ref{eq:suff_condition2}) provide two lower bounds on the period of the Sensor task, i.e., $T_S$. Based on this, to ensure timing constraints are met, $T_S$ needs to satisfy
\begin{equation}
\label{eq:TsBound}
T_S \geq \max(C_M+C_S, \frac{T_{\mathit{tdma}}+C_M+C_S}{N})
\end{equation}

This relation provides a limit on $T_S$, or equivalently, on the  sampling rate, which fulfills the requirements.

\section{Model Checking Approach}\label{sec::modelchecking}
Real-time Maude \cite{DBLP:journals/entcs/OlveczkyM05} and Timed Automata \cite{DBLP:journals/tcs/AlurD94} are two widely used modeling languages which are equipped with model checking facilities. Real-time Maude is a high level declarative programming language supporting specification of real-time and hybrid systems in timed rewriting logic. A network of timed automata models the behavior of timed systems using a set of automata that is equipped with the set of clock variables, which their values are increased in the same rate to model progress in time. Although either of real-time Maude and timed automata can be used as the modeling formalism of the WSAN framework, their faithfulness \ref{DBLP:conf/birthday/Sirjani18} to the WSAN framework is poor. This results in increasing the cost of modeling and model checking. To avoid this inefficiency, we used Timed Rebeca~\cite{DBLP:journals/scp/ReynissonSACJIS14} as the modeling language and benefit from Afra, its model checking toolset \cite{DBLP:conf/facs2/KhamespanahSVK15,DBLP:journals/scp/KhamespanahKS18}. Timed Rebeca is an extension on Rebeca~\cite{DBLP:journals/fuin/SirjaniMSB04,DBLP:conf/birthday/SirjaniJ11} modeling language with time features for modeling and verification of time-critical systems. Rebeca is an actor-based language for modeling concurrent and reactive systems with asynchronous message passing. This characteristics conforms the characteristics of the WSAN framework which consists of a set of concurrently executing components which are communicating by asynchronous message passing. Note that in the development of the Timed Rebeca model of the WSAN framework we do our best to consider the same level of details in comparison with the analytical model. This is necessary to be able to fairly compare the results of the two approaches.

We illustrate the Timed Rebeca language with the simplified model of the WSAN framework, shown in Listing~\ref{fig::simplified}. A Timed Rebeca model consists of a set of reactive classes and the main block. In the main block, actors which are instances of the reactive classes are declared (five actors in this model which are \texttt{medium},  \texttt{cpu}, \texttt{misc}, \texttt{sensorNodeSenderDevice}, and \texttt{receiver}). The body of the reactive class includes the declaration of its known rebecs (line 3), state variables (line 4), and message servers (lines 9 to 12). A message server declaration consists of specifying its signature and a body which contains its corresponding Rebeca statements. The statements in the body can be assignments, conditional statements, enumerated loops, non-deterministic assignment, and method calls. Method calls are sending asynchronous messages to other actors (or to itself in line 11). A modeler can express progress in time using the \texttt{delay} function and communication delay by associating \texttt{after} to method calls.

\begin{lstlisting}[language=rebeca, caption=The Simplified Timed Rebeca implementation of the WSAN framework, label=fig::simplified]
env int samplingRate = 25; // Hz
reactiveclass Sensor(2) {
  knownrebecs { CPU cpu; }
  statevars { int period; }
  Sensor() { 
    self.sensorLoop();
    period = 1000 / samplingRate;
  }
  msgsrv sensorLoop() {
    cpu.sensorEvent() deadline(period);
    self.sensorLoop() after(period);
  }
}
reactiveclass CPU(10) { ... }
reactiveclass Misc(2) { ... }
reactiveclass WirelessMedium(5) { ... }
reactiveclass CommunicationDevice (10) { ... }
main {
  WirelessMedium medium():();
  CPU cpu (sensorNodeSenderDevice, receiver, sensor):();
  Sensor sensor(cpu):();
  Misc misc(cpu):();
  CommunicationDevice sensorNodeSenderDevice(medium):(1);
  CommunicationDevice receiver(medium):(0);
}
\end{lstlisting}

\subsection{Addressing Requirement~\ref{req::inter}}
The first requirement of this model is implemented in the model of \texttt{Sensor} in Listing~\ref{fig::simplified}. The behavior of \texttt{Sensor} is to acquire data and send it to \texttt{CPU} periodically, which is implemented using the message server \texttt{sensorLoop} (lines 10 and 11). Note that the timing values in this model are exact values as they are specified as period of events. To address Requirement~\ref{req::inter} we have to make sure that the sent data is served before the start time of the next period of data acquisition, which is specified by the value of \texttt{period} as the parameter of \texttt{deadline} in line 10. A similar approach is used in the implementation of the \texttt{Miscellaneous} reactive class.

\subsection{Addressing Requirement~\ref{req::intra}}\label{sec::model-checking-intra}

To address the second requirement of the WSAN framework, we have to add the remaining part of the model and put a constraint in a message server which handles sending data to the radio communication device. 


The behavior of \texttt{CPU} as the target of messages of \texttt{Sensor} and \texttt{Misc} is more complicated (Listing~\ref{fig::cpu-model}). Upon receiving a \texttt{miscEvent}, \texttt{CPU} has to represent computation cycles consumed by miscellaneous tasks which is implemented by waiting nondeterministically for one to ten units of time (Note that this value is represented by $C_M$ in the analytical approach in the previous sections). Similarly, after receiving the \texttt{sensorEvent} message from \texttt{Sensor}, \texttt{CPU} waits nondeterministically for one or two units of time, which is the required computation timed for the intra-node data processing. The processed data has to be packed in a packet which has the capacity of storing \texttt{bufferSize} number of data (Note that this value is represented by $N$ in the analytical approach in the previous sections). When the threshold of number of data in a packet is reached (line 17), \texttt{CPU} asks \texttt{senderDevice}, to send the collected data as one packet (line 18). The type of \texttt{senderDevice} is \texttt{RCD} which is the short name for the radio communication device actor. In the model of Listing~\ref{fig::cpu-model} the required computation time of \texttt{sensorEvent} and \texttt{miscEvent} are implemented by nondeterministic values as the worst case computation time is 2 and 10. So, one of the values from 1 to those worst case values may needed at each round of execution, which are modeled by nondeterministic expressions. Note that putting the worst case execution time as the delay values of Listing~\ref{fig::cpu-model} may result in incorrect analysis results as the model may have timing anomaly. As shown in \cite{DBLP:conf/rtss/LundqvistS99}, timing analysis methods which assume that considering the worst-case computation times necessarily corresponds to the worst-case timing behavior of the system does not work correctly for all types of applications and shorter computation times may result in the worst-case timing behavior. 

\begin{lstlisting}[language=rebeca, caption=The Timed Rebeca implementation of \texttt{CPU} reactive class, label=fig::cpu-model]
env int bufferSize = 3; // samples

reactiveclass CPU(10) {
  knownrebecs {RCD senderDevice, receiverDevice;}
  statevars { int collectedSamplesCounter; }

  CPU() { collectedSamplesCounter = 0; }

  msgsrv miscEvent() {
    //Worst case execution time is 10
    delay(?(1, 2, 3, 4, 5, 6, 7, 8, 9, 10));
  }
  msgsrv sensorEvent() {
    //Worst case execution time is 2
    delay(?(1, 2));
    collectedSamplesCounter += 1;
    if (collectedSamplesCounter == bufferSize) {
      senderDevice.send(receiverDevice, 1);
      collectedSamplesCounter = 0;
    }
  }
}
\end{lstlisting}

To fulfill Requirement~\ref{req::intra}, we have to make sure that in the case of sending a \texttt{senderDevice} message in line 18, there is no ongoing sending data in \texttt{RCD}. As we show in  Listing~\ref{fig::communication-device} of  Appendix~\ref{sec::detailed-model}, this requirement is implemented in the body of the \texttt{senderDevice} message server of \texttt{RCD}. Developing the radio communication device actor requires that the wireless communication medium Ether be specified and a communication protocol is implemented. The detailed implementation of Ether and TDMA protocol in \texttt{RCD} are presented in Appendix~\ref{sec::detailed-model}.

\section{Experimental Results and Discussion}\label{sec::comp}
For the aim of better understanding of trade-offs of the approaches, comparison between the analytical analysis and model checking in calculating the maximum sampling rate of the sensor is presented in this section. We defined a set of configurations and prepare the analysis result of each approach on it and provide a discussion on different types of criteria.

\subsection{Results of Applying the Analysis Approaches}
\label{sec:results}
For the case of analytical approach, we consider the following values for the parameters of the model: $T_{\mathit{tdma}}=10$ ms, $C_M=10$ ms, $T_M=120$ ms. Then, we evaluate the model for different values for the WCET of $\tau_S$ and the internal buffer size ($N$), as $C_S \in \{2, 10, 20, 30 \}$ms and $N \in \{1, \ldots, 10\}$. As a result, according to equation (\ref{eq:TsBound}) we obtain the minimum feasible sampling periods, i.e., $T_S$, as shown in Table~\ref{tab:TS} and the maximum sampling rates as shown in Table~\ref{tab:rate}. These tables also contain the minimum sampling periods and maximum sampling rates which are computed using the model checking approach. For this case, we model check the Timed Rebeca model using the same set of configurations and valuation. Comparing the values of these tables shows that in all cases, the maximum possible sampling rates which are calculated by the model checking approach are higher than that of the analytical analysis approach. So, it seems that the analytical analysis approach is a more conservative approach and computes sampling rates based on more pessimistic assumptions.

\begin{table}[!htb]
   \begin{minipage}{0.48\textwidth}
     \centering
     \caption{The minimum feasible sampling periods (milliseconds), computed by both approaches in different configurations}
     \label{tab:TS}
     \begin{tabular}{ll|*{4}{p{0.55cm}}|*{4}{p{0.55cm}}}
       \multicolumn{2}{c}{} & \multicolumn{8}{c}{WCET of $\tau_S$ ($C_S$)} \\ 
       \multicolumn{2}{c}{} & \multicolumn{4}{c}{Analytical} & \multicolumn{4}{c}{Model Checking}\\ 
& & 2    & 10   & 20   & 30 & 2 & 10 & 20 & 30   \\
        \hline
\multirow{10}{*}{\rotatebox[origin=c]{90}{Internal Buffer Size (N)}} 
& 1  & 20   & 20   & 30   & 40  & 11 & 11   & 22 & 33   \\
& 2  & 12   & 20   & 30   & 40  & 11 & 11   & 22 & 33   \\
& 3  & 12   & 20   & 30   & 40  & 11 & 11   & 22 & 33   \\
& 4  & 12   & 20   & 30   & 40  & 11 & 11   & 22 & 33   \\
& 5  & 12   & 20   & 30   & 40  & 11 & 11   & 22 & 33   \\
& 6  & 12   & 20   & 30   & 40  & 11 & 11   & 22 & 33   \\
& 7  & 12   & 20   & 30   & 40  & 11 & 11   & 22 & 33   \\
& 8  & 12   & 20   & 30   & 40  & 11 & 11   & 22 & 33   \\
& 9  & 12   & 20   & 30   & 40  & 11 & 11   & 22 & 33   \\
& 10 & 12   & 20   & 30   & 40  & 11 & 11   & 22 & 33 \\ \bottomrule
     \end{tabular}
   \end{minipage}\hfill
   \begin {minipage}{0.48\textwidth}
     \centering
     \caption{The maximum sampling rates (samples per second), computed by both approaches in different configurations}
     \label{tab:rate}
     \begin{tabular}{ll|*{4}{p{0.55cm}}|*{4}{p{0.55cm}}}
       \multicolumn{2}{c}{} & \multicolumn{8}{c}{WCET of $\tau_S$ ($C_S$)} \\ 
       \multicolumn{2}{c}{} & \multicolumn{4}{c}{Analytical} & \multicolumn{4}{c}{Model Checking}\\ 
& & 2    & 10   & 20   & 30 & 2 & 10 & 20 & 30   \\

        \hline
\multirow{10}{*}{\rotatebox[origin=c]{90}{Internal Buffer Size (N)}} 
& 1  & 50   & 50   & 33 & 25 & 90 & 90   & 45 & 30   \\
& 2  & 83 & 50   & 33 & 25 & 90 & 90   & 45 & 30   \\
& 3  & 83 & 50   & 33 & 25 & 90 & 90   & 45 & 30   \\
& 4  & 83 & 50   & 33 & 25 & 90 & 90   & 45 & 30   \\
& 5  & 83 & 50   & 33 & 25 & 90 & 90   & 45 & 30   \\
& 6  & 83 & 50   & 33 & 25 & 90 & 90   & 45 & 30   \\
& 7  & 83 & 50   & 33 & 25 & 90 & 90   & 45 & 30   \\
& 8  & 83 & 50   & 33 & 25 & 90 & 90   & 45 & 30   \\
& 9  & 83 & 50   & 33 & 25 & 90 & 90   & 45 & 30   \\
& 10 & 83 & 50   & 33 & 25 & 90 & 90   & 45 & 30   \\ \bottomrule
     \end{tabular}
   \end{minipage}
\end{table}

\subsection{Discussion}
Different modeling approaches affect the complexity and flexibility of the modeling process and the models. It can also affect the analyzability and precision of the results
\cite{DBLP:conf/birthday/Sirjani18}. In the following, we discuss three criteria which help in comparing the two approaches: (i) flexibility and extensibility, i.e., the  effort needed for modifying and extending a model, (ii) accuracy of the results, and (iii) the resource consumption for the analysis process, which concerns how much computation power is needed for the approach.

\subsubsection{Flexibility and Extensibility.}
For comparing the analytical  and  model checking approaches, from the flexibility and extensibility point of view, we study how changing the communication protocol of the model from TDMA to B-MAC is realized in two approaches. 
Changing the communication protocol does not affect the task model and resource model which are introduced in Section~\ref{sec::analytical}. But, the algorithm of using the resource (the communication medium as a critical section) has to be modified. Based on the model of this paper, the behavior of communication protocols is abstracted to a value which shows the worst-case delay between requesting for a data transmission to its corresponding successful data transmission. Thus, changing the communication protocol only changes this value. 
In order to show this, we have modified the analysis to accommodate B-MAC communication protocol by using the schedulability analysis of B-MAC given in~\cite{Jung_JCN12}. 

According to the analysis, the communication delay of a sender node is composed by several parameters, including (i) the initial backoff time $t_{b1}$; (ii) the frozen time between every sequential retransmission $t_{f1}$; (iii) the congestion backoff time $t_{b2}$; and (iv) the frozen time during the congestion backoff $t_{f2}$. The congestion can occur multiple times in a network, thus the times for them should be accounted as well. The packet transmission time ($t_{pkt}$) should be also added to the communication delay. Eq (\ref{eq:bmac}) calculates the communication delay of a sender node in B-MAC, denoted by $t_{sd}$~\cite{Jung_JCN12}, where $k$ is the maximum number of retransmission in the case of congestion. This equation is representative of B-MAC transmission delay, thus $T_\mathit{tdma}$ in Eq~(\ref{eq:TsBound}) must be replaced with $t_{sd}$.

\begin{equation}
\label{eq:bmac}
t_{sd} = t_{b1} + t_{f1} + k(t_{b2} + t_{f2}) + T_{pkt}
\end{equation}

Considering the same experimental setup as in Section~\ref{sec:results}, we compute the minimum sampling periods and maximum sampling rate of sensors. 
Note that in B-MAC communication we assumed the maximum number of retransmission for a successful transmission is 4 (the maximum number of sender nodes in the sample network).

Although changing the communication protocol looks minor and easy to apply in Eq (\ref{eq:bmac}), the  worst-case delay is computed based on an assumption on the maximum number of retransmissions which is not easy to compute accurately. The accurate timing behavior of B-MAC protocol is illustrated in collaboration between network nodes in runtime. So, a very complicated formula is needed to express the accurate value of the worst-case delay in B-MAC, or it has to be approximated to a value which corresponds to the very pessimistic configuration of the application (for some cases it is infinity), which is unreal for the majority of the cases.

In contrast, using Timed Rebeca, no expertise in finding delays is needed and the protocol has to be modeled using the language statements. Using B-MAC protocol, a radio communication device tries to detect when the channel is free and sends data upon receiving a request from CPU. The detailed implementation of this protocol and how collisions are handled are depicted in Appendix~\ref{sec::BMAC}, 

Changing the behavioral type of sensor or miscellaneous tasks is another modification which can be used to compare flexibility in two approaches. Assume that we want to modify the models in a way that miscellaneous task is no more periodic but aperiodic. In this case, all the formulas which are used in Section~\ref{sec::analytical} to specify the task model have to be modified. As we will discuss in the following subsection, the current formulation of the model does not support non-periodic tasks and new modeling formalism has to be used to address aperiodic tasks. In contrast, the only modification needed in the Timed Rebeca model is in the statements of the \texttt{miscLoop} message server. 

\subsubsection{Results Accuracy.}
As shown in the experimental result, the model checking approach provided better (more accurate) results. This is because our analytical approach exhibits an approximate solution. One of the main sources of needs for such approximation is in the low expressiveness of the chosen task model (our modeling was not exact). 

To make the result of the analytical approach more accurate, we have to change the task model to a model with a higher level of expressiveness. In this paper, we employed a periodic task model for the analytical approach. As a result, we could not model the possible initial offset of the activation of the tasks; we assumed a pessimistic case where the tasks are released synchronously while there are task models with a higher level of  expressiveness. This means that we can take into account more details of the system behavior with such task models, possibly resulting in a more accurate solution. 


\begin{figure*}
\centering
\includegraphics[width=0.9\textwidth]{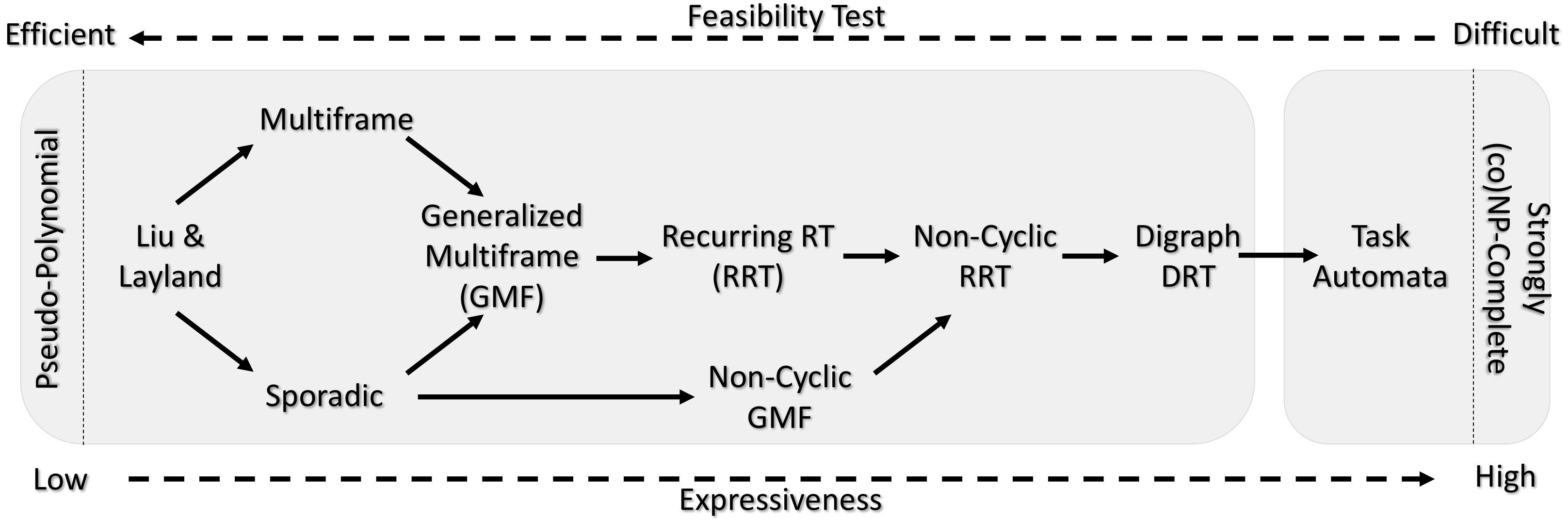}
\caption{A generalization relation between some real-time task models (adopted from~\cite{Stigge2011})}
\label{fig::task-models2}
\end{figure*}

Figure~\ref{fig::task-models2} illustrates a hierarchy of real-time task models with different levels of expressiveness. The model we used for the analysis falls into the category of sporadic tasks. As shown in Figure~\ref{fig::task-models2}, there are more expressive task models which can provide, possibly, more accurate analysis results at the cost of lower efficiency. In particular, the DRT task model, proposed by Stigge et al.~\cite{Stigge2011}, allows non-deterministic initial offset of a task release. 
This is a potential future work to model and analyze the problem with such a task model. In addition, the inherent complexity of the problem enforces some relaxation in the model, results in pessimistic approximations. Fersman et al.~\cite{Fersman2007} have shown that the feasibility problem is undecidable if the following conditions hold: (1) The scheduling method is preemptive; (2) execution time of the jobs is not fixed, this value is picked from a range; (3) the completion  time of a job can influence the release of a next job. 
On the other hand, if even one of these conditions does not hold, the problem is decidable. Such complexities lead us to use approximation in our analysis. Specifically, to verify Relation~(\ref{eq:sched_condition}), we need a specific response time analysis which can compute (a lower bound on) the difference of the response time of a job and its N$_{th}$ successor. To the best of our knowledge, there is not such analysis for the known real-time task models in the literature. However, to avoid a complex analysis, we replace exact values with lower and upper bounds (Eq.~(\ref{eq:suff_condition})). While this relaxation does not compromise soundness of the analysis, it can lead to pessimistic results. We have made an initial attempt to make use of more expressive task models for which pseudo-polynomial time feasibility tests exists. This appeared that using such models still cannot increase the accuracy of the analysis.
An alternative model to describe and analyze the system under study is Data-flow graph (DFG)~\cite{Lee87}. Although, there is limited work on schedulability analysis of DFGs \cite{singh2018}.

\subsubsection{Effort needed for doing the Analysis.}
As the last criterion, we compare the computational power needed for the two approaches to do the analysis. For the case of analytical approach, the  computation power is negligible as the result is computed by setting parameters of the schedulability formula. This argument is valid for any possible values of parameters. But, in the model checking approach the state space of the model has to be generated and explored. For the case of Timed Rebeca model of this paper, the needed time is 2 seconds and 2039 states are generated on a desktop computer with 1 CPU (2 cores) and 8 GB of RAM storage, running High Sierra OS X 10.13.5, which are not too much resources. However, as mentioned in Section~\ref{sec::model-checking-intra}, we used non-deterministic expressions to address worst-case execution times in \texttt{CPU} and increasing the value of worst-case execution time increases the options of the non-deterministic expression. Increasing the options of non-deterministic expressions may exponentially increase the size of the state space. Therefore, in this sense, model checking approach is less scalable comparing to the analytical approach.

\section{Conclusions and Future work}\label{sec::con}

In this paper, we presented a comparison study between using an analytical analysis approach and a model checking approach on a WSAN application, to obtain the highest possible data acquisition frequency of nodes. Configuring the highest possible frequency in nodes leads to achieve more efficient use of resources and in turn reducing costs. As for the analytical analysis approach we used a response time analysis technique and for the model checking approach we used Timed Rebeca, which is a modeling language for time-critical systems. The comparison study shows that the model checking approach delivers more accurate results compared to the analytical approach for the presented WSAN application. In order to improve the analytical approach, the task model has to be improved to give a better expressiveness and more behavioral details, however at the same time it makes the model more complex to analyze. 
Future work aims at study the effects of other task models, e.g., DRT task model, on the analytical approach compared to model checking approaches.

\section*{Acknowledgements}
The work on this paper has been supported in part by the project "Self-Adaptive Actors: SEADA" (nr. 163205-051) of the Icelandic Research Fund, “Modeling and Analyzing Collaborating Machines: MACMA" from Software Center, Sweden, and "Dependable Platforms for Autonomous Systems and Control: DPAC" from KKS, Sweden.

%
%
%
\bibliographystyle{splncs04}
\bibliography{ref}

\begin{thebibliography}{10}
\providecommand{\url}[1]{#1}
\csname url@samestyle\endcsname
\providecommand{\newblock}{\relax}
\providecommand{\bibinfo}[2]{#2}
\providecommand{\BIBentrySTDinterwordspacing}{\spaceskip=0pt\relax}
\providecommand{\BIBentryALTinterwordstretchfactor}{4}
\providecommand{\BIBentryALTinterwordspacing}{\spaceskip=\fontdimen2\font plus
\BIBentryALTinterwordstretchfactor\fontdimen3\font minus
  \fontdimen4\font\relax}
\providecommand{\BIBforeignlanguage}[2]{{%
\expandafter\ifx\csname l@#1\endcsname\relax
\typeout{** WARNING: IEEEtran.bst: No hyphenation pattern has been}%
\typeout{** loaded for the language `#1'. Using the pattern for}%
\typeout{** the default language instead.}%
\else
\language=\csname l@#1\endcsname
\fi
#2}}
\providecommand{\BIBdecl}{\relax}
\BIBdecl

\bibitem{Linderman2012}
L.~Linderman, K.~Mechitov, and B.~F. Spencer, ``{TinyOS-Based Real-Time
  Wireless Data Acquisition Framework for Structural Health Monitoring and
  Control},'' \emph{Structural Control and Health Monitoring}, 2012.

\bibitem{sha2004real}
L.~Sha, T.~Abdelzaher, K.-E. {\AA}rz{\'e}n, A.~Cervin, T.~Baker, A.~Burns,
  G.~Buttazzo, M.~Caccamo, J.~Lehoczky, and A.~K. Mok, ``Real time scheduling
  theory: A historical perspective,'' \emph{Real-time systems}, 2004.

\bibitem{tinyos}
J.~Hill, R.~Szewczyk, A.~Woo, S.~Hollar, D.~Culler, and K.~Pister, ``System
  architecture directions for networked sensors,'' \emph{SIGPLAN Notices},
  2000.

\bibitem{Liu_JACM73}
C.~Liu and J.~Layland, ``Scheduling algorithms for multiprogramming in a
  hard-real-time environment,'' \emph{Journal of the Association for Computing
  Machinery}, 1973.

\bibitem{Joseph_CJ86}
M.~Joseph and P.~Pandya, ``Finding response times in a real-time system,''
  \emph{The Computer Journal}, 1986.

\bibitem{Thiele_IEEE00}
L.~Thiele, S.~Chakraborty, and M.~Naedele, ``Real-time calculus for scheduling
  hard real-time systems,'' in \emph{International Symposium on Circuits and
  Systems. Emerging Technologies for the 21st Century}, 2000.

\bibitem{Waszniowski2008}
L.~Waszniowski and Z.~Hanz{\'a}lek, ``Formal verification of multitasking
  applications based on timed automata model,'' \emph{Real-Time Systems}, 2008.

\bibitem{Yi_tool04}
P.~Kr{\v{c}}{\'a}l and W.~Yi, ``Decidable and undecidable problems in
  schedulability analysis using timed automata,'' in \emph{Tools and Algorithms
  for the Construction and Analysis of Systems}, 2004.

\bibitem{Xu_GCC13}
X.~Xu, X.~Y. Li, and M.~Song, ``Distributed scheduling for real-time data
  collection in wireless sensor networks,'' in \emph{IEEE Global Communications
  Conference}, 2013.

\bibitem{Xia_IEEE17}
C.~Xia, X.~Jin, L.~Kong, and P.~Zeng, ``Bounding the demand of
  mixed-criticality industrial wireless sensor networks,'' \emph{IEEE Access},
  2017.

\bibitem{Cheng_TOR11}
B.~C. Cheng, H.~H. Yeh, and P.~H. Hsu, ``Schedulability analysis for hard
  network lifetime wireless sensor networks with high energy first
  clustering,'' \emph{IEEE Transactions on Reliability}, 2011.

\bibitem{Furfaro_MCSIT07}
A.~Furfaro and L.~Nigro, ``Modelling and schedulability analysis of real-time
  sequence patterns using {Time Petri Nets and Uppaal},'' in
  \emph{International Multiconference on Computer Science and Information
  Technology}, 2007.

\bibitem{Xu_IEEE02}
D.~Xu, X.~He, and Y.~Deng, ``Compositional schedulability analysis of real-time
  systems using {Time Petri Nets},'' \emph{IEEE Transactions on Software
  Engineering}, 2002.

\bibitem{Fersman_tools02}
E.~Fersman, P.~Pettersson, and W.~Yi, ``Timed automata with asynchronous
  processes: Schedulability and decidability,'' in \emph{Tools and Algorithms
  for the Construction and Analysis of Systems}, 2002.

\bibitem{Li_IPDPSW10}
G.~Li, X.~Cai, and S.~Yuen, ``Modeling and analysis of real -time systems with
  mutex components,'' in \emph{International Symposium on Parallel Distributed
  Processing, Workshops and Phd Forum}, 2010.

\bibitem{Fersman_JC04}
E.~Fersman and W.~Yi, ``A generic approach to schedulability analysis of
  real-time tasks,'' \emph{Nordic Journal of Computing}, 2004.

\bibitem{Sheng_WMTV10}
W.~Sheng, Y.~Gao, L.~Xi, and X.~Zhou, ``Schedulability analysis for multicore
  global scheduling with model checking,'' in \emph{International Workshop on
  Microprocessor Test and Verification}, 2010.

\bibitem{Guan_ISORC08}
N.~Guan, Z.~Gu, M.~Lv, Q.~Deng, and G.~Yu, ``Schedulability analysis of global
  fixed-priority or {EDF} multiprocessor scheduling with symbolic
  model-checking,'' in \emph{International Symposium on Object and
  Component-Oriented Real-Time Distributed Computing}, 2008.

\bibitem{spencer2015}
B.~F. Spencer, H.~Jo, K.~A. Mechitov, J.~Li, S.-H. Sim, R.~E. Kim, S.~Cho,
  L.~E. Linderman, P.~Moinzadeh, R.~K. Giles \emph{et~al.}, ``Recent advances
  in wireless smart sensors for multi-scale monitoring and control of civil
  infrastructure,'' \emph{Journal of Civil Structural Health Monitoring},
  vol.~6, no.~1, pp. 17--41, 2016.

\bibitem{DBLP:conf/iscc/El-Hoiydi02}
A.~El{-}Hoiydi, ``Spatial {TDMA} and {CSMA} with preamble sampling for low
  power ad hoc wireless sensor networks,'' in \emph{Symposium on Computers and
  Communications}, 2002.

\bibitem{DBLP:conf/sensys/PolastreHC04}
J.~Polastre, J.~L. Hill, and D.~E. Culler, ``Versatile low power media access
  for wireless sensor networks,'' in \emph{International Conference on Embedded
  Networked Sensor Systems}, 2004.

\bibitem{Davis_11}
R.~Davis, S.~Kollmann, V.~Pollex, and F.~Slomka, ``{Controller Area Network
  (CAN) schedulability analysis with FIFO queues},'' in \emph{Euromicro
  Conference on Real-Time Systems}, 2011.

\bibitem{DBLP:journals/entcs/OlveczkyM05}
P.~C. {\"O}lveczky and J.~Meseguer, ``{R}eal-{T}ime {M}aude 2.1,''
  \emph{Electr. Notes Theor. Comput. Sci.}, vol. 117, pp. 285--314, 2005.

\bibitem{DBLP:journals/tcs/AlurD94}
R.~Alur and D.~L. Dill, ``{A} {T}heory of {T}imed {A}utomata,'' \emph{Theor.
  Comput. Sci.}, vol. 126, no.~2, pp. 183--235, 1994.

\bibitem{DBLP:conf/birthday/Sirjani18}
\BIBentryALTinterwordspacing
M.~Sirjani, ``Power is overrated, go for friendliness! expressiveness,
  faithfulness, and usability in modeling: The actor experience,'' in
  \emph{Principles of Modeling - Essays Dedicated to Edward A. Lee on the
  Occasion of His 60th Birthday}, ser. Lecture Notes in Computer Science,
  M.~Lohstroh, P.~Derler, and M.~Sirjani, Eds., vol. 10760.\hskip 1em plus
  0.5em minus 0.4em\relax Springer, 2018, pp. 423--448. [Online]. Available:
  \url{https://doi.org/10.1007/978-3-319-95246-8\_25}
\BIBentrySTDinterwordspacing

\bibitem{DBLP:journals/scp/ReynissonSACJIS14}
A.~H. Reynisson, M.~Sirjani, L.~Aceto, M.~Cimini, A.~Jafari,
  A.~Ing{\'{o}}lfsd{\'{o}}ttir, and S.~H. Sigurdarson, ``Modelling and
  simulation of asynchronous real-time systems using timed {Rebeca},''
  \emph{Science of Computer Programming}, 2014.

\bibitem{DBLP:conf/facs2/KhamespanahSVK15}
E.~Khamespanah, M.~Sirjani, M.~Viswanathan, and R.~Khosravi, ``Floating time
  transition system: More efficient analysis of timed actors,'' in
  \emph{International Conference in Formal Aspects of Component Software},
  2015.

\bibitem{DBLP:journals/scp/KhamespanahKS18}
E.~Khamespanah, R.~Khosravi, and M.~Sirjani, ``An efficient {TCTL} model
  checking algorithm and a reduction technique for verification of timed actor
  models,'' \emph{Science of Computer Programming}, 2018.

\bibitem{DBLP:journals/fuin/SirjaniMSB04}
M.~Sirjani, A.~Movaghar, A.~Shali, and F.~S. de~Boer, ``{M}odeling and
  {V}erification of {R}eactive {S}ystems using {R}ebeca,'' \emph{Fundamental
  Information}, 2004.

\bibitem{DBLP:conf/birthday/SirjaniJ11}
M.~Sirjani and M.~M. Jaghoori, ``{T}en {Y}ears of {A}nalyzing {A}ctors:
  {R}ebeca {E}xperience,'' in \emph{Formal Modeling: Actors, Open Systems,
  Biological Systems}, 2011.

\bibitem{DBLP:conf/rtss/LundqvistS99}
T.~Lundqvist and P.~Stenstr{\"{o}}m, ``Timing anomalies in dynamically
  scheduled microprocessors,'' in \emph{Real-Time Systems Symposium}, 1999.

\bibitem{Jung_JCN12}
S.~Jung, N.~Choi, and T.~Kwon, ``An iterative analysis of single-hop b-mac
  networks under poisson traffic,'' \emph{Journal of Communications and
  Networks}, February 2012.

\bibitem{Stigge2011}
M.~Stigge, P.~Ekberg, N.~Guan, and W.~Yi, ``The digraph real-time task model,''
  in \emph{Real-Time and Embedded Technology and Applications Symposium}, 2011,
  pp. 71--80.

\bibitem{Fersman2007}
E.~Fersman, P.~Krcal, P.~Pettersson, and W.~Yi, ``Task automata:
  Schedulability, decidability and undecidability,'' \emph{Information and
  Computation}, 2007.

\bibitem{rebeca}
``{Rebeca Formal Modeling Language},'' \url{http://www.rebeca-lang.org/}.

\end{thebibliography}
\newpage
\appendix
\section{Timed Rebeca Model of Ether and TDMA Protocol}\label{sec::detailed-model}
The reactive class of \texttt{Ether} (Listing~\ref{fig::communication-medium}) has three message servers: these are responsible for sending the status of the medium, broadcasting data, and resetting the status of the medium after a successful transmission. Broadcasting data takes place by sending data to an \texttt{RCD} which results in setting the values of \texttt{senderDevice} and \texttt{receiverDevice} to their corresponding actors. So, the status of \texttt{Ether} can be easily examined by the value of \texttt{receiverDevice} (i.e., using \texttt{null} as the value of \texttt{receiverDevice} is interpreted as medium is free, line 7). This way, as shown in lines 22 and 23, upon successfully data transmission, the value of \texttt{receiverDev} and \texttt{senderDev} have be set to \texttt{null} to show that the transmission is completed. The main behavior of \texttt{Ether} is data broadcasting which is implemented in lines 9 to 20. Before the start of broadcasting, the \texttt{Ether} status is checked (line 11) and data-collision error is raised in the case of trying for more that one simultaneous data broadcasts (line 18). With a successful data broadcast, \texttt{Ether} sends an acknowledgment to itself (line 14) and the sender (line 15), and informs the receiver of the number of packets sent to it (line 16).

\begin{lstlisting}[language=rebeca,caption=The Timed Rebeca implementation of \texttt{Ether} reactive class, label=fig::communication-medium]
reactiveclass Ether(5) {
  statevars { RCD senderDev, receiverDev; }

  Ether() {
    senderDev = null;
    receiverDev = null;
  }  
  msgsrv getStatus() { ((RCD)sender).receiveStatus(receiverDev != null); }
  msgsrv broadcast(RCD receiver, int packets) {
    byte OnePacketTT = ?(5, 6, 7); // ms(transmission time)
    if(senderDev == null) {
      senderDev = (RCD)sender;
      receiverDev = receiver;
      self.broadcastingIsCompleted() after(packets * OnePacketTT);
      ((RCD)sender).receiveResult(true) after(packets * OnePacketTT);
      receiver.receiveData(receiver, packets);
    } else {
      ((RCD)sender).receiveResult(false);
    }
  }
  msgsrv broadcastingIsCompleted() {
    senderDev = null;
    receiverDev = null;
  }
}
\end{lstlisting}

The Timed Rebeca implementation of the TDMA protocol in \texttt{RCD} is depicted in Listing~\ref{fig::communication-device}. It shows that sending a packer, the \texttt{send} message server of \texttt{RCD} has to be called. As shown in line 18 and 19, upon receiving a request for sending data, \texttt{receiverDevice} and \texttt{sendingData} are set to the input parameters and data transmission is started. These state variables are set to null values upon finishing the transmission. So, to fulfill Requirement~\ref{req::intra}, we have to make sure that \texttt{receiverDevice} is set to \texttt{null}, i.e. there is no ongoing sending data, as implemented in line 17.

The TDMA protocol defines a cycle, over which each node in the network has one chances (a time slot) to transmit a packet or a series of packets. If a node has data available to transmit during its alloted time slot, it may be sent immediately. Otherwise, packet sending is delayed until reaching the next time slot. In the Timed Rebeca model of Listing~\ref{fig::communication-device} the periodic behavior of TDMA slot is implemented in the body of \texttt{handleTDMASlot} message server. As depicted in line 23, the value of \texttt{inActivePeriod} is toggled to show that whether the node is in its associated time slot. Upon starting an associated time slot of node, the existing pending data are sent (line 27) a \texttt{handleTDMASlot} is scheduled for terminating the time slot (line 28). Using this implementation, when \texttt{CPU} sends a packet to \texttt{RCD}, the packet is appended to the list of pending packets which are waiting for the next time slot of the node. For the sake of simplicity, some details of \texttt{RCD} are omitted in Listing~\ref{fig::communication-device}. The complete source code of this model is available on the Rebeca homepage~\cite{rebeca}.


\begin{lstlisting}[language=rebeca, caption=The Timed Rebeca implementation of the TDMA protocol in \texttt{RCD}, label=fig::communication-device]
reactiveclass RCD (10) {
  knownrebecs { Ether medium; }
  statevars {
    int id, slotSize, sendingData;
    boolean busyWithSending, inActivePeriod;
    RCD receiverDevice;
  }

  RCD(byte myId) {
    id = myId;
    inActivePeriod = false;
    sendingData = 0;
    receiverDevice = null;
    ...
  }
  msgsrv send(RCD receiver, int data) {
    assertion(receiverDevice == null);
    receiverDevice = receiver;
    sendingData = data;
    self.checkPendingData();
  }
  msgsrv handleTDMASlot() {
    inActivePeriod = !inActivePeriod;
    if(inActivePeriod) {
      int remainedTime = tmdaSlotSize - currentMessageWaitingTime;
      assertion(remainedTime > 0);
      self.checkPendingData();
      self.handleTDMASlot() after(remainedTime);
    } else {
      self.handleTDMASlot() after((slotSize * (numberOfNodes - 1))- currentMessageWaitingTime);
    }
  }
  ...
}
\end{lstlisting}

\section{Timed Rebeca Model of B-MAC Protocol}\label{sec::BMAC}
As shown in Listing~\ref{fig::bmac-communication-device}, using B-MAC protocol, a radio communication device tries to detect when the channel is free and sends data upon receiving a request from CPU (line 19 of Listing~\ref{fig::bmac-communication-device}). In the case of busy channel, retrying for transmission takes place in line 20. If the channel is free, the requested packet is sent immediately (line 22) regardless of the status of the other nodes of the network. This way, collisions may occur. To resolve collision, the \texttt{receiveResult} message server has been changed to be aware of the result of sent data. In the case of collision, the state of the transmitter is reset and sending process is started from the beginning (line 32). In comparison with the TDMA protocol, B-MAC protocol does not need complicated and expensive synchronization methods. It also avoids data fragmentation. So, it would be more inefficient to coordinate long messages and B-MAC expects short messages, which is common for the passing packets of WSAN applications.

\begin{lstlisting}[language=rebeca, caption=The Timed Rebeca implementation of B-MAC protocol in \texttt{RCD}, label=fig::bmac-communication-device]
reactiveclass RCD (10) {
  knownrebecs { WirelessMedium medium; }
  statevars {
    int id, sendingData;
    RCD receiverDevice;
  }
  RCD(int myId) {
    id = myId;
    sendingData = 0;
    receiverDevice = null;
  }
  msgsrv send(RCD receiver, int data, int packetsNumber) {
    assertion(receiverDevice == null);
    receiverDevice = receiver;
    sendingData = data;
    medium.getStatus();
  }
  msgsrv receiveStatus(boolean result) {
    if (!result) {
      medium.getStatus() after(OnePacketTT);
    } else {
      medium.broadcast(receiverDevice, sendingData, packetsNumber);
      delay(OnePacketTT * packetsNumber);
    }
  }
  msgsrv receiveResult(boolean result) {
    if (result) {
      sendingPacketsNumber = 0;
      receiverDevice = null;
      sendingData = 0;
    } else {
      medium.getStatus() after (OnePacketTT);
    }
  }
  msgsrv receiveData(RCD receiver, int data, int recPacketsNumber) { ... }
}
\end{lstlisting}

\end{document}